\documentclass[preprint,12pt,3p]{elsarticle}

\usepackage{amsmath,amssymb,amsfonts}
\usepackage{mathrsfs}
\usepackage{chemformula}
\usepackage{color}
\usepackage[normalem]{ulem}
\usepackage{stmaryrd}
\usepackage{diagbox}
\usepackage[parfill]{parskip}

 \biboptions{sort&compress}

\newcommand{\blue}[1]{\textcolor{black}{#1}}

\newcommand{\beq}{\begin{equation}}
\newcommand{\eeq}{\end{equation}}
\newcommand{\pfr}[2]{\ensuremath{\frac{\partial #1}{\partial #2}}}

\newcommand{\ep}{\epsilon}

\newcommand\Pec{\mbox{\textit{Pe}}}

\newcommand\Lew{\mbox{\textit{Le}}}
\newcommand\Sto{\mbox{\textit{St}}}
\newcommand\Rey{\mbox{\textit{Re}}}
\newcommand{\vect}[1]{\mathbf{#1}}

\journal{Combustion and Flame}

\begin{document}

\begin{frontmatter}

\title{\blue{Effective Lewis number and burning speed for flames propagating} in small-scale spatio-temporal periodic flows}

\author{Prabakaran Rajamanickam, Joel Daou}
\address{Department of Mathematics, University of Manchester, Manchester M13 9PL, UK}

\begin{abstract}
 Propagation of premixed flames having thick reaction zones in rapidly-varying, small-scale, zero-mean, spatio-temporal periodic flows is considered. Techniques of large activation energy asymptotics and homogenization theory are used to determine the effective Lewis number $\Lew_{\mathrm{eff}}$ and the effective burning speed ratio $S_T/S_L$, which are influenced by the flow through flow-enhanced diffusion. The resultant effective diffusivity matrix is, in general, neither a scalar nor a diagonal matrix and therefore induces anisotropic effects on the propagation of multi-dimensional flames. As the flow Peclet number $\Pec$ becomes large, the flow-enhanced fuel diffusion coefficient and the thermal diffusivity behave respectively like $(\Pec\Lew)^\sigma$ and $\Pec^\sigma$, where $\Lew$ is the Lewis number and $\sigma\leq 2$ is a constant which depends on the flow and the direction of flame propagation. The maximal value $\sigma=2$ is achieved for steady, unidirectional, spatially periodic shear flows, while for steady two-dimensional square vortices, we have $\sigma=1/2$. In general, the constant $\sigma$ is determined by solving a linear partial differential equation. The scaling laws for the diffusion coefficients lead to corresponding scaling laws for the effective Lewis number and the effective burning speed ratio of the form $\Lew_{\mathrm{eff}}\simeq \Lew^{1-\sigma}$ and $S_T/S_L \sim (\Pec/\Lew)^{\sigma/2}$. Effects of thermal expansion and volumetric heat loss on the flame are also briefly discussed. In particular, it is shown that the quenching limit is enlarged by a factor $1/\Lew^\sigma$ for $\Lew<1$ and diminished by the same factor for $\Lew>1$, due to the flow-enhanced diffusion. The potential implications of the results to better understand turbulent combustion are discussed. A special emphasis is placed on the dependence of the flame on $\Lew$ in the presence of high-intensity, small-scale flows. In particular, it is shown that this dependence is intimately linked to the flow through Taylor-dispersion like enhanced diffusion, rather than through the traditional molecular diffusion coupled with curvature effects. The flow-dependent effective Lewis number identified may also provide an explanation to the peculiar experimental observation that turbulence appears to facilitate ignition in $\Lew>1$ mixtures and to inhibit it in $\Lew<1$ mixtures.
\end{abstract}


\end{frontmatter}

\section*{Novelty and significance statement}
An original study, combining asymptotic analysis and homogenization theory, is applied to describe flame propagation in small-scale, spatio-temporal periodic flow fields. Scaling laws are derived for the effective burning speed and the effective Lewis number for high-intensity small-scale flows, which are useful to better understand the behaviour of turbulent premixed flames in the distributed reaction zone regime. The formula for the flow-dependent effective Lewis number identified herein may explain the peculiar experimental observation that turbulence appears to facilitate ignition in $\Lew>1$ mixtures and to inhibit it in $\Lew<1$ mixtures. The high-intensity small-scale flows are shown to increase the quenching limit due to volumetric heat losses in $\Lew<1$ mixtures and decrease it in $\Lew>1$ mixtures. 

\section*{Author contributions}
PR carried out the research and wrote the paper. JD designed the research and wrote the paper.

\section{Introduction}

In this paper, we study the problem of flame propagation in spatially or more generally spatio-temporally periodic flows having single length and time scales. A number of theoretical~\cite{ashurst1991short,berestycki1991flame,brailovsky1995extinction,audoly2000reaction} and computational~\cite{aldredge1996premixed,kagan1998flame,kagan2000flame,kagan2002activation,kagan2005effect,kadowaki2009dynamic} investigations have been devoted to this problem in the past and more recently in~\cite{kagan2020flame,feng2021influence}. Experimental studies have also addressed this problem, notably in order to gain insight into flame propagation in the presence of Taylor-Couette vortices~\cite{shy1992experimental,zhu1994simulation,vaezi2000laminar}. One of the main motivations of these and similar studies has been to improve our understanding of premixed turbulent combustion. \blue{The reader is referred to specialized reviews such as \cite{peters1941turbulent,lipatnikov2005molecular} for an overview of the main issues  in the  vast field of turbulent combustion. Here, we simply note that} most theoretical works have focused primarily on the thin flame or thin reaction zone regimes, describing flame propagation in a large-scale flow field. On the other hand, the effect of small-scale flows is ubiquitous in turbulent combustion, notably in the distributed reaction zone regime~\cite{peters1941turbulent,williams2000progress} where some flow scales can become smaller than the size of the reaction zone. 

The current paper focuses on the thick reaction-zone limit~\cite{rajamanickam2023thick} to elucidate the influence of small-scale periodic flows on flame propagation.
\blue{Particular attention is devoted to characterising the effective Lewis number and burning speed in such flows. The focus of the study is partly motivated by apparent disagreement revived recently regarding the effective Lewis number, $\Lew_{\rm{eff}}$ in stronlgy turbulent flows. Specifically, whereas according to common views~\cite{peters1941turbulent,aspden2011turbulence,lapointe2015differential,aspden2019towards}, $\Lew_{\rm{eff}}$ should be unity in such conditions, recent studies~\cite{rieth2022enhanced,lee2021influence,lee2022lewis,cai2022turbulent} suggest otherwise. Notably, it is argued in~\cite{rieth2022enhanced} that the molecular Lewis number effects are still important in strongly turbulent flows and are most active at scales small compared with the flame thickness. Although our laminar periodic flow model cannot fully settle the disagreement regarding $\Lew_{\rm{eff}}$ in turbulent combustion, it can provide a helpful insight by determining $\Lew_{\rm{eff}}$ for the small-scale laminar flows considered. This} problem is treated analytically in the limit of large activation energy, with the reaction zone thickness being larger than the flow length scale.


The paper is organized is as follows. The characteristic scales involved in thick reaction-zone flames propagating in small-scale periodic flows are introduced in $\S$\ref{sec:scal}. The problem governing equations and boundary conditions are then formulated in $\S$\ref{sec:gov}, and these are the basis of an asymptotic analysis carried out in $\S$\ref{sec:asym}. The scaling laws for the effective burning speed and the effective Lewis number are obtained in $\S$\ref{sec:largePe} for large values of the Peclet number. The results are illustrated for two classes of prototypical flow fields, namely for unsteady unidirectional flows and for the so-called Childress-Soward flows which are steady  and two dimensional. Potential implications for premixed turbulent combustion are discussed in $\S$\ref{sec:turb}. The results of $\S$\ref{sec:largePe}  are complemented by selected illustrative results for arbitrary Peclet numbers in $\S$\ref{sec:arbPe}. Further extensions of the work including the effect of thermal expansion, heat loss are briefly discussed in $\S$\ref{sec:th} and $\S$\ref{sec:fut} and followed by conclusions in~$\S$\ref{sec:con}.

\section{Scalings for thick reaction-zone flames}
\label{sec:scal}

Consider in a reactive mixture a spatio-temporal periodic flow field $\vect v(\vect x,t)$, where $\vect x$ is the dimensional position vector and $t$ the dimensional time. Let the characteristic flow amplitude, spatial period  and temporal period be, $U$, $l_{\rm{cell}}$ and $t_{\rm{cell}}$, respectively.  Further, let us also assume that the mean value of $\vect v(\vect x,t)$ is zero in a suitable frame of reference. Then, the heat transport process may be characterised by two dimensionless numbers, namely, the Peclet number $\Pec$ and the Stokes number $\Sto$, defined by
\begin{equation}
    \Pec = \frac{Ul_{\rm{cell}}}{D_T}, \qquad \qquad  \Sto =  \frac{l_{\rm{cell}}^2/D_T}{t_{\rm{cell}}}, \nonumber
\end{equation}
where $D_T$ denotes the thermal diffusivity of the gas mixture. 

\begin{figure}[h!]
\centering
\includegraphics[scale=0.7]{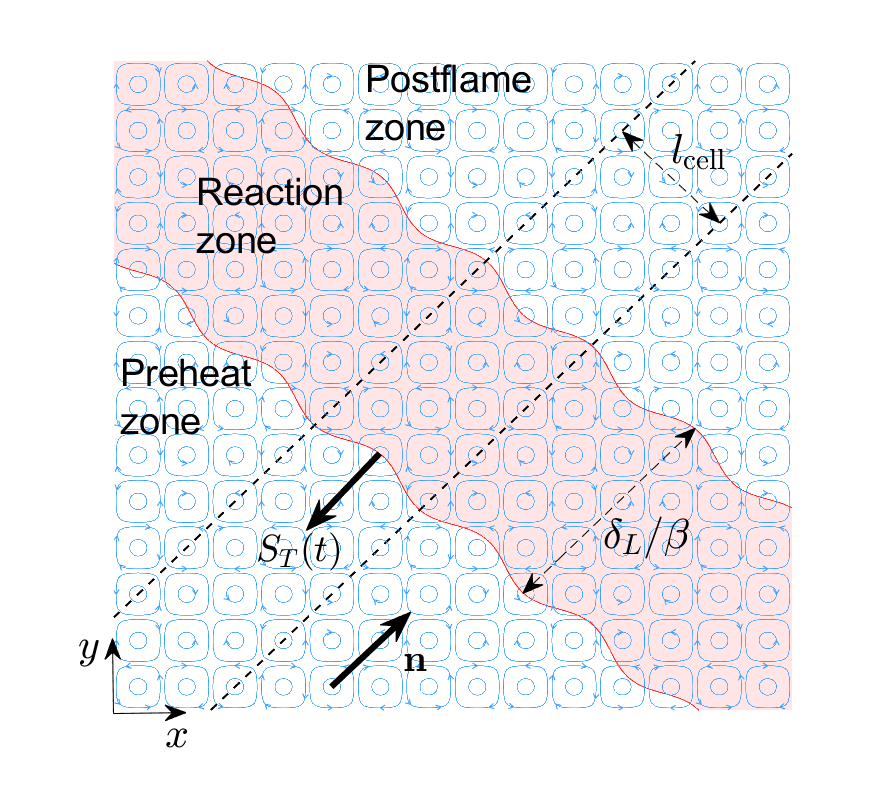}
\caption{Schematic illustration of the flame structure in the thick reaction-zone limit, $\delta_L/\beta \sim l_{\rm{cell}}$~\cite{rajamanickam2023thick}. The flame is assumed to be periodic (or independent of) directions perpendicular to $\vect n$. Here, we shall consider the ultra-thick reaction-zone regime wherein $\delta_L/\beta \gg l_{\rm{cell}}$.} \label{fig:sch}
\end{figure}

The thermal and chemical properties of the reactive mixture define a laminar burning speed $S_L$ and a laminar flame thickness $\delta_L=D_T/S_L$. The thickness of the reaction zone is then given by $\delta_L/\beta$, where $\beta$ is the Zeldovich number; these quantities are defined below in~\eqref{zel}. By comparing the flow scale with the reaction-zone thickness, combustion modes can be classified into three regimes~\cite{rajamanickam2023thick}, namely a thin reaction-zone  ($\delta_L/\beta\ll l_{\rm{cell}}$), a thick reaction-zone ($\delta_L/\beta\sim l_{\rm{cell}}$) and an ultra-thick reaction-zone ($\delta_L/\beta\gg l_{\rm{cell}}$) regimes. The thin reaction-zone regime includes both thin flames ($\delta_L\ll l_{\rm{cell}}$) and moderately thick flames ($\delta_L\sim l_{\rm{cell}}$)~\cite{daou2002thick}. A schematic illustration of the flame structure in the thick reaction-zone limit is shown in Fig.~\ref{fig:sch}. In the current paper, we shall focus on the ultra-thick reaction-zone regime where $\delta_L\gg \delta_L/\beta \gg l_{\rm{cell}}$.

Assume that the flame structure propagates in the periodic flow field in a definite direction, say $-\vect n$, with some propagation speed. This is justified if the structure is periodic in (or, independent of) directions perpendicular to $\vect n$, as we shall assume. Furthermore, since we consider zero-mean flows, the flame propagation speed is also the effective burning speed $S_T(t)$, in the first approximation. Specifically, the function $S_T(t)$ (in general, periodic in $t$) can be defined as the total instantaneous burning rate per unit cross-sectional area normal to $\vect n$ of the infinite strip depicted in Fig.~\ref{fig:sch}. Numerical computations of flame propagation are typically performed only for such infinitely long strips~\cite{aldredge1996premixed,kagan2000flame,kagan2002activation,kagan2005effect}. 

As mentioned above, the present study deals with the limit $\delta_L\gg \delta_L/\beta \gg l_{\rm{cell}}$. This requirement implies that for fixed Peclet number
\begin{equation}
    \frac{S_T}{U}\sim \frac{S_L}{U} = \frac{\ep}{\Pec} \ll 1 \quad \text{since} \quad \ep \equiv \frac{l_{\rm{cell}}}{\delta_L}\ll 1.   \nonumber
\end{equation}
 That is to say, the burning speed is small when compared to the flow amplitude, as it is the case for sufficiently thick flames. In addition, in the case of time-dependent flows, the homogenization analysis below requires that $ \delta_L/\beta S_L \gg t_{\rm{cell}}$, where $\delta_L/\beta S_L$ is the residence time in the reaction zone. 

 Under the assumptions above, the flame can be regarded in the first approximation as being planar on the flame (large) scale $\vect x \sim \delta_L$, whereas it will be non-planar on the flow (small) scale $\vect x \sim l_{\rm{cell}}$. The burning speed $S_T$ will be a time-independent constant in the first approximation and involve corrections of order $\ep$ in the following approximation.

The effect of a rapidly-varying small-scale motion is known to manifest as a diffusion process on the large scale. The homogenization theory can be used to quantify the flow-enhanced diffusion analytically by taking advantage of the separation of scales between the flame and the flow. An excellent review of this technique, applied to a non-reactive scalar field, has been provided by Majda and Kramer~\cite{majda1999simplified}. We shall employ this technique to our flame propagation problem, following~\cite{majda1999simplified} closely.

\section{Governing equations}
\label{sec:gov}

 It is advantageous to \blue{adopt} a reference frame that is moving with the flame. The time and space coordinates are non-dimensionalized using the flow scales, 
\begin{equation}
    \tau = \frac{tU}{l_{\mathrm{cell}}}, \qquad  \boldsymbol{\xi} = \frac{1}{l_{\rm{cell}}}\left(\vect x + \vect n\int_0^t S_T\,dt\right)=\frac{\vect x}{l_{\rm{cell}}} + \frac{\epsilon\vect n}{\Pec}\int_0^\tau S\, d\tau,  \label{nondim}
\end{equation}
where $S \equiv S_T/S_L$ is the ratio of effective burning speed to the laminar flame speed. Therefore $(\boldsymbol{\xi},\tau)$ are appropriate independent variables for the small-scale flow field, but not for the thick flame. From a small-scale viewpoint, the coordinate shift $\boldsymbol{\xi}-\vect x/l_{\rm{cell}}$ between the laboratory frame and the flame-fixed frame is negligible at leading order according to~\eqref{nondim} since $\ep \ll 1$. In the frame attached to the flame, the laboratory-frame vector field $\vect v(\vect x/l_{\mathrm{cell}}, t/t_{\mathrm{cell}})$ transforms into  
 \begin{equation}
     \vect u(\boldsymbol{\xi},\tau) \equiv \frac{1}{U}\vect v\left(\frac{\vect x}{l_{\rm{cell}}}(\boldsymbol{\xi},\tau),\frac{\tau\Sto}{\Pec}\right)   \nonumber
 \end{equation}
 after scaling with $U$. In the limit $\ep\to 0$, the right-hand side of this equation may be expanded in a Taylor series as
 \begin{equation}
     \vect u(\boldsymbol{\xi},\tau) = \frac{\vect v}{U} - \frac{\ep S_0 \tau }{U\Pec} \vect n \cdot \nabla_\xi\vect v + \cdots
 \end{equation}
where $\vect v,\nabla_\xi\vect v,\dots$ are evaluated at $(\boldsymbol{\xi},\tau)$ and $S_0$ denotes the leading-order of $S$.  
 

The unburnt reacting mixture is assumed to be fuel lean, whose combustion chemistry is modeled by a single-step irreversible Arrhenius reaction with the fuel burning rate (in mass units) per unit volume given by $\rho B Y_F e^{-E/RT}$, that involves the pre-exponential factor $B$, the gas density $\rho$, the fuel mass fraction $Y_F$, the temperature $T$, the activation energy $E$ and the universal gas constant $R$. Further, for simplicity, we shall \blue{adopt} the thermo-diffusive approximation in which density and molecular diffusivities are constant and briefly discuss the effect of variable density later. Also, we \blue{introduce} the Zeldovich number $\beta$, heat release parameter $\alpha$ and the laminar flame speed $S_L$ \blue{(for $\beta\gg 1$)} by
\begin{equation}
    \beta = \frac{E(T_{ad}-T_u)}{RT_{ad}^2}, \quad \alpha = \frac{T_{ad}-T_u}{T_{ad}}, \quad S_L=\left(2\Lew \beta^{-2} B  D_{T} e^{-E/RT_{ad}}\right)^{1/2}. \label{zel}
\end{equation}
In these expressions, $\Lew$ denotes the Lewis number, $T_u$ the unburnt gas temperature and $ T_{ad} = T_u + qY_{F,u}/c_p$  the adiabatic flame temperature, where $q$ is the heat release rate per unit mass of fuel burnt, $Y_{F,u}$ is the fuel mass fraction in the unburnt mixture and $c_p$ is the constant-pressure specific heat. 

The scaled fuel mass fraction and temperature are defined by
\begin{equation}
    y_F = \frac{Y_F}{Y_{F,u}}, \qquad \theta = \frac{T-T_u}{T_{ad}-T_u}.  \nonumber
\end{equation}
The non-dimensional governing equations in a frame attached to the flame are given by
\begin{align}
    \Pec\pfr{y_F}{\tau}  + (\epsilon S\vect n+\Pec \vect u) \cdot \nabla_{\xi} y_F &= \frac{1}{\Lew}\nabla_\xi^2 y_F - \epsilon^2 \omega(y_F,\theta),  \label{yF}\\
 \Pec\pfr{\theta}{\tau} + (\epsilon S\vect n+\Pec \vect u)  \cdot \nabla_{\xi}\theta &= \nabla_\xi^2\theta+ \epsilon^2 \omega(y_F,\theta) \label{theta}
\end{align}
where 
\begin{equation}
    \omega(y_F,\theta) = \frac{\beta^2 y_F}{2\Lew }\exp\left[\frac{\beta(\theta-1)}{1+\alpha(\theta-1)}\right]. \nonumber
\end{equation}
The boundary conditions for $y_F$ and $\theta$ need to be prescribed in terms of the large scale variable $ \vect x/\delta_L$, i.e., $\ep\boldsymbol{\xi}$. They are given by
\begin{align}
     \ep \boldsymbol{\xi}\cdot \vect n \rightarrow -\infty: &\qquad \qquad y_F=1, \quad \theta = 0, \label{BC1}\\
     \ep \boldsymbol{\xi}\cdot \vect n \rightarrow +\infty: &\qquad \qquad y_F=0, \quad \ep^{-1}\nabla_{\xi}\theta\cdot \vect n = 0 \label{BC2}
\end{align}
in the direction of $\vect n$. Periodicity conditions are imposed in other spatial directions and in time.

\section{Asymptotic analysis in the double limit $\ep\rightarrow 0$, $\beta\ep\rightarrow 0$}
\label{sec:asym}
In this section, we carry out an asymptotic analysis of the problem~\eqref{yF}-\eqref{BC2} in the double limit $\ep\to 0$, $\beta\ep\to 0$. At leading order, the flame structure is steady and one dimensional  on the large scale $\vect X \equiv \ep\boldsymbol{\xi}\sim 1$. To describe this structure on this scale, we use the multiple-scale technique, involving the small-scale coordinates $(\boldsymbol{\xi},\tau)$ and the large-scale coordinate $\vect X$. Consequently, derivatives transform according to 
\begin{equation}
      \nabla_{\xi} \rightarrow \nabla_{\xi} + \epsilon \nabla_{X} \nonumber
\end{equation}
where $\nabla_\xi$ and $\nabla_{X}$ are the gradient operators in the small-scale and large-scale coordinates. The appropriate expansion for the solution can be written as
\begin{align}
    y_F &= F_0(\vect X) + \epsilon F_1(\vect X,\boldsymbol{\xi},\tau) + \epsilon^2 F_2(\vect X,\boldsymbol{\xi},\tau) + \cdots, \nonumber \\
    \theta  & = \Theta_0(\vect X) + \epsilon \Theta_1(\vect X,\boldsymbol{\xi},\tau) + \epsilon^2 \Theta_2(\vect X,\boldsymbol{\xi},\tau) + \cdots, \nonumber \\
    S &= S_0 + \epsilon S_1(\tau) + \epsilon^2 S_2(\tau) + \cdots. \nonumber
\end{align}

Substituting these expansions into~\eqref{yF}-\eqref{BC2} and collecting terms of different orders of $\ep$, we obtain a series of equations for $F_i$, $\Theta_i$ and $S_i$. The equations that arise at leading order are identically satisfied as we have already anticipated that $F_0$ and $\Theta_0$ are independent of small-scale variables. The equations at the next two orders are found to be
\begin{align}
    \mathcal{L}_F F_1 = &-  \Pec \Lew\,\vect v\cdot \nabla_X F_0,  \label{FF1}\\ 
    \mathcal{L}_T \Theta_1 =&- \Pec\, \vect v\cdot \nabla_X \Theta_0,   \label{TT1} \\
    \mathcal{L}_F F_2 = &- S_0\Lew\left[ \vect n-\tau\vect n \cdot \nabla_\xi\vect v\right]\cdot (\nabla_XF_0 + \nabla_{\xi} F_1)   -  \Pec \Lew\,\vect v\cdot\nabla_X F_1   + \nabla_{X}^2F_0  \nonumber \\ & +2 \nabla_{\xi}\cdot  \nabla_{X} F_1-\Lew\,\omega(F_0,\Theta_0), \label{FF2} \\
    \mathcal{L}_T \Theta_2 = &- S_0\left[ \vect n-\tau\vect n \cdot \nabla_\xi\vect v\right]\cdot (\nabla_X\Theta_0 + \nabla_{\xi} \Theta_1)   - \Pec\, \vect v\cdot\nabla_X \Theta_1  + \nabla_{X}^2\Theta_0  \nonumber \\ &+2 \nabla_{\xi}\cdot  \nabla_{X} \Theta_1+\omega(F_0,\Theta_0)  \label{TT2}
\end{align}
where
\begin{align}
    \mathcal{L}_F \equiv \Pec \Lew\, \pfr{}{\tau} +  \Pec \Lew\,\vect v \cdot \nabla_{\xi} -\nabla_\xi^2, \qquad
    \mathcal{L}_T \equiv  \Pec\, \pfr{}{\tau} + \Pec\,  \vect v \cdot \nabla_{\xi} -  \nabla_\xi^2 \label{LFLT}
\end{align}
are differential operators that act on small-scale variables.

Solutions for the first-order equations~\eqref{FF1}-\eqref{TT1} are obtained by assuming 
\begin{align}
    F_1 &= F_{1a}(\vect X) +  \Pec \Lew\,\vect K_F\cdot \nabla_X F_0 , \label{F1}\\
    \Theta_1 &= \Theta_{1a}(\vect X) + \Pec \, \vect K_T \cdot \nabla_X\Theta_0 \label{T1}
\end{align}
where the vectors $\vect K_F(\boldsymbol{\xi},\tau)$ and $\vect K_T(\boldsymbol{\xi},\tau)$ are the periodic solutions of 
\begin{equation}
    \mathcal{L}_F\vect K_F = - \vect v ,\qquad \qquad \mathcal{L}_T \vect K_T = -  \vect v.  \label{KFKT}
\end{equation}
Since $\vect v$ is of the form $\vect v = \vect v(\boldsymbol{\xi},\tau \Sto/\Pec)$, the function $\vect K_F$ depends only on the two parameters $\Pec\Lew$ and $\Sto\Lew$ while, similarly, $\vect K_T$ only depends on  $\Pec$ and $\Sto$.

The second-order non-homogeneous equations~\eqref{FF2}-\eqref{TT2} governing $F_2$ and $\Theta_2$ are solvable only if the right-hand sides satisfy a solvability condition. Specifically, following~\cite{majda1999simplified}, the solvability condition states that given a periodic function $f(\boldsymbol{\xi},\tau)$, the equation $\mathcal{L}_F g(\boldsymbol{\xi},\tau) = f(\boldsymbol{\xi},\tau)$ has a smooth periodic solution if and only if $f(\boldsymbol{\xi},\tau)$ has zero mean. Imposing this condition on equation~\eqref{TT2}, for instance,  we obtain
\begin{align}
    \left \langle- S_0\left[ \vect n-\tau\vect n \cdot \nabla_\xi\vect v\right]\cdot (\nabla_X\Theta_0 + \nabla_{\xi} \Theta_1)   -  \Pec \,\vect v\cdot\nabla_X \Theta_1   + \nabla_{X}^2\Theta_0 +2 \nabla_{\xi}\cdot  \nabla_{X} \Theta_1 +\omega\right \rangle =0 \nonumber 
\end{align}
where $\langle\,\cdot\,\rangle $ denotes an average\footnote{The average is defined by $\langle \varphi \rangle \equiv \frac{\Sto}{\Pec}\int_{0}^{\Pec/\Sto}\int_0^1\int_0^1\int_0^1 \varphi \,d\xi_1d\xi_2d\xi_3d\tau$.} over the small scale variables $\boldsymbol{\xi}$ and $\tau$. 

Clearly, terms such as $ - S_0 \vect n \cdot \nabla_X\Theta_0 + \nabla_{X}^2\Theta_0 + \omega$ that do not depend on the small-scale variables are unaffected by the averaging operation. On the other hand, all terms that contain $\nabla_{\xi}$ can be shown, using the divergence theorem and the periodicity boundary condition, to vanish identically. The only remaining contribution is due to $\Pec \,\vect v\cdot\nabla_X \Theta_1$ in which only the second term in~\eqref{T1} survives upon averaging. This particular contribution $\Pec\langle \vect v\cdot\nabla_X \Theta_1\rangle= \Pec^2\langle \vect v\cdot\nabla_X (\vect K_T\cdot \nabla_X\Theta_0)\rangle$ can be written as $\Pec^2 \nabla_X \cdot (\langle\vect v\vect K_T\rangle\cdot \nabla_X \Theta_0)$ or, equivalently $\Pec^2 \langle\vect v\vect K_T\rangle : \nabla_X \nabla_X \Theta_0$ since $\vect v$ is independent of $\vect X$ and the averaging operation affects only the small-scale variables. Furthermore since the Hessian matrix $\nabla_X\nabla_X$ which represents the tensor ${\partial^2}/{\partial_{X_i}\partial_{X_j}}$ is clearly symmetric, we may symmetrise the product $\vect v \vect K_T$.

Now, on applying the solvability conditions for both dependent variables and simplifying the results, we obtain
\begin{align}
    S_0  \vect n \cdot \nabla_X F_0 &= \frac{1}{\Lew} \nabla_X \cdot (\boldsymbol{\mathcal{D}}_F\cdot\nabla_X F_0)  - \omega(F_0,\Theta_0), \label{F0}\\
    S_0  \vect n \cdot \nabla_X \Theta_0 &=  \nabla_X \cdot (\boldsymbol{\mathcal{D}}_T\cdot\nabla_X \Theta_0)  + \omega(F_0,\Theta_0) \label{T0}
\end{align}
where the effective diffusion matrices $ \boldsymbol{\mathcal{D}}_F= \boldsymbol{\mathcal{D}}_F(\Pec\Lew,\Sto\Lew)$ and $ \boldsymbol{\mathcal{D}}_T= \boldsymbol{\mathcal{D}}_F(\Pec,\Sto)$ are given\footnote{The equality of the second relation to the first is shown readily in index notation~\cite{majda1999simplified}: substitute $v_i = \mathcal{L}_F K_{F,i}$ to obtain $\frac{1}{2}  \langle v_iK_{F,j} + v_jK_{F,i} \rangle= \frac{1}{2}  \langle K_{F,j}\mathcal{L}_F K_{F,i} + K_{F,i}\mathcal{L}_F K_{F,j} \rangle$, which can be re-written, using~\eqref{LFLT}, as $\frac{1}{2}  \langle \mathcal{L}_F (K_{F,i}K_{F,j}) + 2\nabla_{\xi} K_{F,i}\cdot \nabla_{\xi} K_{F,j}\rangle$ and then impose the solvability condition for the first term $\langle \mathcal{L}_F (K_{F,i}K_{F,j})\rangle=0$.}  by
\begin{align}
    \boldsymbol{\mathcal{D}}_F &= \vect I - \frac{1}{2}\Pec^2\Lew^2\left\langle \vect v\vect K_F  +  (\vect v\vect K_F )^T\right\rangle = \vect I + \Pec^2\Lew^2\left\langle \nabla_\xi\vect K_F\circ (\nabla_\xi\vect K_F)^T\right\rangle \label{DF} \\
    \boldsymbol{\mathcal{D}}_T &= \vect I - \frac{1}{2}\Pec^2\left\langle \vect v\vect K_T  +  (\vect v\vect K_T )^T\right\rangle = \vect I + \Pec^2\left\langle \nabla_\xi\vect K_T\circ (\nabla_\xi\vect K_T)^T\right\rangle \label{DT}
\end{align}
in which $\vect I$ denotes the identity matrix and the symbol $\circ$ represents element-wise matrix multiplication such that, for example, $D_{T,ij}= \delta_{ij} + \Pec^2 \langle\nabla_{\xi} K_{T,i}\cdot \nabla_{\xi} K_{T,j}\rangle$.

While formulas~\eqref{DF}-\eqref{DT} for the effective diffusion matrices are of general use in a variety of problems such as problems involving propagation and stability, they are more transparent if an axis of the coordinate system is chosen to be along $\vect n$. Let $\boldsymbol{\mathcal{R}}$ be the rotation matrix which transforms the original coordinate vector $\vect X$ into a new coordinate vector $\vect X'=\boldsymbol{\mathcal{R}}\vect X$ such that the $X_1'$-axis is directed along $\vect n$. In the rotated coordinate system, $F_0(X_1')$ and $\Theta_0(X_1')$ are function only of $X_1'$ and therefore equations~\eqref{F0}-\eqref{T0} simplify to
\begin{align}
    S_0 \frac{d F_0}{dX_1'} &= \frac{\mathcal{D}_{F,11}'}{\Lew} \frac{d^2F_0}{dX_1'^2}  - \omega(F_0,\Theta_0), \label{FF0}\\
    S_0  \frac{d \Theta_0}{dX_1'} &=  \mathcal{D}_{T,11}' \frac{d^2\Theta_0}{dX_1'^2}  + \omega(F_0,\Theta_0)\label{TT0}
\end{align}
where $\boldsymbol{\mathcal{D}}_F' = \boldsymbol{\mathcal{R }}\boldsymbol{\mathcal{D}}_F  \boldsymbol{\mathcal{R}}^T$ and $\boldsymbol{\mathcal{D}}_T' = \boldsymbol{\mathcal{R }}\boldsymbol{\mathcal{D}}_T  \boldsymbol{\mathcal{R}}^T$. Also, the boundary conditions~\eqref{BC1}-\eqref{BC2} reduce to
\begin{equation}
     X_1' \rightarrow -\infty: \,\,\,  y_F-1=\theta = 0 \quad \text{and} \quad
     X_1' \rightarrow +\infty: \,\,\, y_F= \frac{d\Theta_0}{dX_1'} = 0. \label{BCC}
\end{equation}

The solution of problem~\eqref{FF0}-\eqref{BCC} for large $\beta$, as done~e.g. in~\cite{daou2018taylor},  provides the following formulas for the effective burning speed and the effective Lewis number
\begin{equation}
   \frac{S_T}{S_L}\simeq S_0 = \frac{\mathcal{D}_{T,11}'}{\sqrt{\mathcal{D}_{F,11}'}}= \sqrt{\frac{\Lew_{\mathrm{eff}}}{\Lew}\mathcal{D}_{T,11}'} \quad\quad \text{and} \quad\quad \Lew_{\rm{eff}} = \Lew \frac{\mathcal{D}_{T,11}'}{\mathcal{D}_{F,11}'}. \label{S0}
\end{equation}
 Note that the dependence of $S_0$ on $\mathcal{D}_{T,11}'$ and $\mathcal{D}_{F,11}'$ is intuitively correct as it extends the dependence of the laminar speed $S_L$ on $D_T$ and $D_F$ in~\eqref{zel}, namely $S_L \propto D_T/\sqrt{D_F}$, by replacing $D_T$ and $D_F$ with corresponding flow-enhanced values.

\section{Scaling laws for large Peclet numbers}
\label{sec:largePe}

The dependence of the effective burning speed ratio $S_T/S_L$ and the effective Lewis number $\Lew_{\rm{eff}}$ on $\Pec$ is determined by the enhanced diffusion coefficients, as indicated in~\eqref{S0}. For large values of $\Pec$,  we may assume that the asymptotic behaviour of the effective diffusion coefficients is of the form
\begin{equation}
   \mathcal{D}_{F,11}'-1 \sim (\Pec \Lew)^\sigma, \quad \mathcal{D}_{T,11}'-1 \sim \Pec^\sigma \quad \text{for} \quad \Pec\gg 1,  \label{sigma}
\end{equation}
and similar behaviours (with different exponents $\sigma$) for other elements of the matrix $\boldsymbol{\mathcal{D}}_F-\vect I$. Then~\eqref{S0} implies that 
\begin{equation}
    \Lew_{\rm{eff}} \simeq \Lew^{1-\sigma}, \qquad  \qquad  \frac{S_T}{S_L} \sim \left(\frac{\Pec}{\Lew}\right)^{\sigma/2},   \label{LeS0}
\end{equation}
provided the exponent $\sigma$, which needs to be computed as done below, is positive. Negative values of $\sigma$ indicate that there is no enhancement of diffusion with respect to molecular diffusion for $\Pec \gg 1$ and therefore 
\begin{equation}
    \Lew_{\rm{eff}} =\Lew, \qquad  \qquad  \frac{S_T}{S_L} =1.  \nonumber
\end{equation}
 In fact, as demonstrated in~\cite{majda1999simplified}, $\sigma$ is bounded from above, namely $ \sigma \leq 2$. This upper bound indicates that maximal enhancement of diffusion is achieved when $\sigma=2$.  The maximal enhancement in fact occurs for steady, unidirectional periodic (or confined) shear flows and is associated with the Taylor's dispersion mechanism~\cite{daou2018taylor}. It is instructive to consider two classes of flow fields, namely, unsteady unidirectional flows for which the exponent $\sigma$ can be determined explicitly and the two-dimensional so-called Childress-Soward flows for which $\sigma$ will be computed numerically.

\subsection{Unsteady unidirectional flows}

Diffusion enhancement has been studied in the context of unsteady unidirectional periodic shear flows
by Zeldovich~\cite{zeldovich1982exact}, who provided an exact solution for the effective diffusion coefficient. The similar diffusion problem in confined geometries has been studied by Watson~\cite{watson1983diffusion}. The general unidirectional shear flow periodic in time and space with zero mean  may be written in the form of a double Fourier series as
\begin{equation}
    v_1(\xi_2,\tau) = \sum_{(k,m)\neq (0,0)} \hat v_{k,m} e^{2\pi i(k\xi_2 + m\tau St/Pe)}, \qquad v_2=v_3=0. \nonumber
\end{equation}
Using this expression in~\eqref{KFKT} to determine $\vect K_F$ and then using~\eqref{KFKT} to determine $\boldsymbol{\mathcal{D}}_F'$, we find that the only non-zero element of $\boldsymbol{\mathcal{D}}_F'-\mathbf{I}$ is $\mathcal{D}_{F,11}' - 1$. This term is given by
\begin{equation}
     \mathcal{D}_{F,11}' - 1 = \sum_{(k,m)\neq (0,0)} \frac{(k\Pec\Lew)^2 |\hat v_{k,m}|^2}{4\pi^2 k^4 + (m\Sto\Lew)^2}, \label{zeldo}
\end{equation}
 a result which is equivalent to formula (55) of~\cite{majda1999simplified}, which generalizes an earlier formula originally derived by Zeldovich~\cite{zeldovich1982exact}. It is worth noting that when $\Sto\sim \Pec$, i.e., when the time scale $t_{\mathrm{cell}}$ corresponds to $l_{\mathrm{cell}}/U$, $\mathcal{D}_{F,11}' - 1$ tends to a constant independent of $\Pec$ as $\Pec\to \infty$. This indicates that the exponent $\sigma=0$ and that the enhancement of diffusion remains bounded as noted by Zeldovich. On the other hand, when $\Sto=0$ corresponding to a steady flow, or more generally when $\Sto \ll \Pec$ corresponding to a quasi-steady flow, formula~\eqref{zeldo} indicates that $\mathcal{D}_{F,11}' - 1\sim (\Pec\Lew)^2$, which is the maximal enhancement aforementioned.  For $\Sto \gg \Pec$, diffusion enhancement is negligible according to~\eqref{zeldo}. 

\subsection{Steady two-dimensional Childress-Soward flows}

\begin{figure}[h!]
\centering
 \advance\leftskip0.1cm
\hspace*{-1.2in}
\includegraphics[scale=0.53]{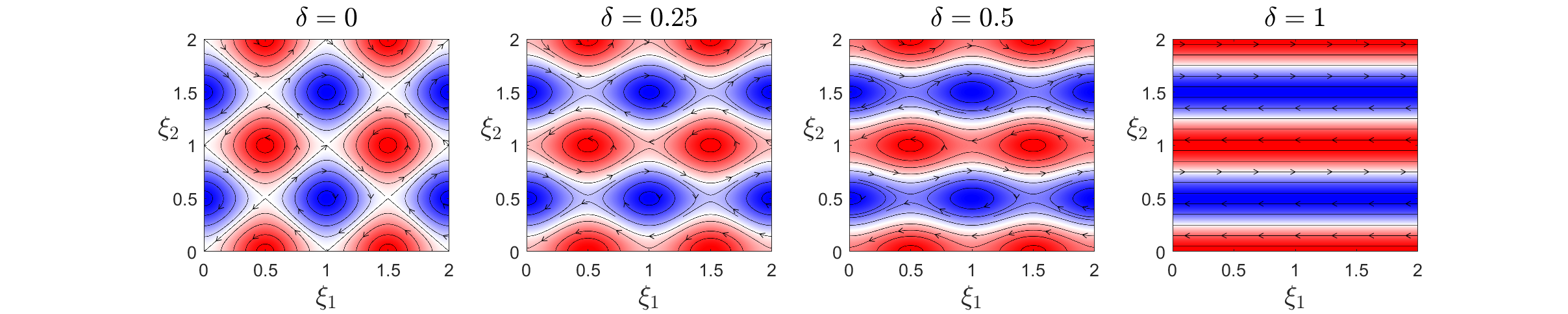}
\caption{Streamline plots for the Childress-Soward flows~\eqref{child} for selected values of $\delta$. The colours describe the scalar vorticity field (normalized by its maximum value with red indicating a positive or counter-clockwise vorticity and blue a negative or clockwise vorticity).} \label{fig:cs}
\end{figure}

In addition to the unidirectional flows, another prototypical flow which has been used in theoretical studies such as~\cite{ashurst1991short,berestycki1991flame,brailovsky1995extinction,audoly2000reaction} on flame-flow interaction is the so-called vortical (or cellular flow). A useful class of simple steady flows depending on a parameter $0\leq\delta\leq 1$ which encompasses both the shear and cellular flows is the so-called Childress-Soward flows~\cite{childress1989scalar}. The velocity components of the Childress-Soward flow in a suitable coordinate system are given by
\begin{equation}
    v_1 = -(1+\delta) \sin(2\pi \xi_2), \quad v_2 = -(1-\delta) \sin(2\pi \xi_1), \quad v_3 = 0. \label{child}
\end{equation}
 As shown by the streamline plots in Fig.~\ref{fig:cs}, we have a cellular flow consisting of square vortices for $\delta=0$ and a unidirectional shear flow directed along $\xi_1$-axis for $\delta=1$, with intermediate values of $\delta$ representing a series of cats-eye vortices with varying degrees of eddy-like/shear motion.

For the Childress-Soward flows (in the frame of reference chosen) given by~\eqref{child}, $ \boldsymbol{\mathcal{D}}_T' - \vect I$ is a diagonal matrix. Corresponding to these flows, we now calculate numerically the exponent $\sigma$ for each diagonal entry of the matrix $ \boldsymbol{\mathcal{D}}_T' - \vect I$. For ${\mathcal{D}}_{T,11}' - 1$ we first solve~\eqref{KFKT} for $\vect K_T$, then evaluate $\boldsymbol{\mathcal{D}}_T'$ using~\eqref{DT} and then finally fit the data for large $\Pec$ to the profile
\begin{equation}
   {\mathcal{D}}_{T,11}' - 1 \sim \Pec^\sigma.
\end{equation}
We proceed similarly to determine the exponent $\sigma$ for ${\mathcal{D}}_{T,22}' - 1$. \blue{The numerical results are obtained by solving the inhomogeneous elliptic PDEs~\eqref{KFKT} using COMSOL Multiphysics software. The equations are solved subject to periodic boundary conditions along with an additional condition, say $\vect K_T(0,0)=0$. The latter condition is needed since the solution is unique only within an additive constant.}  For the empirical fit, we have used the numerical results corresponding to $\Pec$ in the range $[6\times10^3,10^4]$. \blue{It should be cautioned that the accuracy in determining the exponent $\sigma$ depends on its value; the larger the value of $\sigma$, the better is the fitting accuracy.}

 \begin{figure}[h!]
\centering
\includegraphics[scale=0.7]{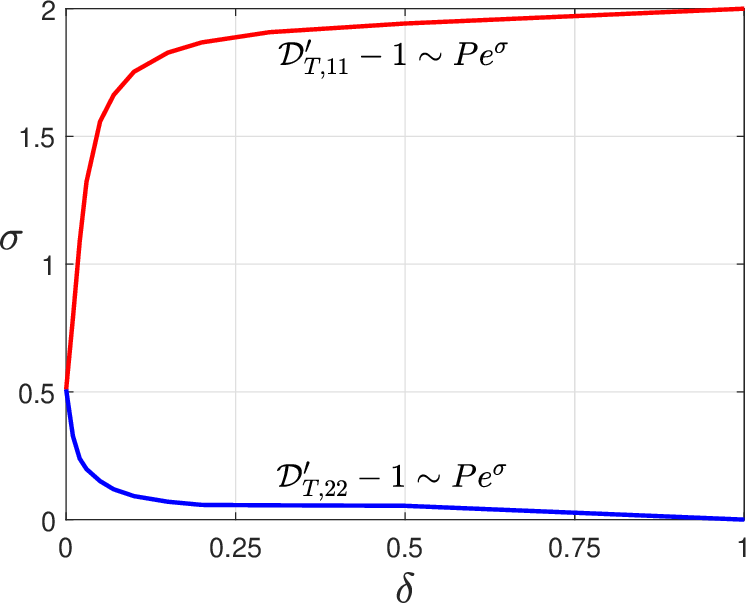}
\caption{The exponents $\sigma$ in the formula $ {\mathcal{D}}_{T,11}' - 1 \sim \Pec^\sigma$  (red line) and in the formula $ {\mathcal{D}}_{T,22}' - 1 \sim \Pec^\sigma$ (blue line) vs. the parameter $\delta$ appearing in~\eqref{child}, obtained by fitting numerical computations for $\Pec$ in the range $[6\times10^3,10^4]$.} \label{fig:sig}
\end{figure}

The \blue{computed} exponents $\sigma$ corresponding to the two diagonal elements are plotted in Fig.~\ref{fig:sig} as a function of $\delta$. It can be observed from the figure that when $\delta=0$, $\sigma= 1/2$ in agreement with the predictions of past investigations~\cite{childress1979alpha,fannjiang1994convection} on square vortices. Similarly for $\delta=1$ corresponding to a unidirectional shear flow, it is seen that the enhancement of diffusion is in the $\xi_1$-direction only, in line with the conclusions of the previous subsection. 

We note that as $\delta$ is increased, the enhancement of diffusion increases in the $\xi_1$-direction and decreases in the other direction. This translates into the effective diffusion process becoming more anisotropic as $\delta$ is increased. For $\delta=0$, there is no anisotropy in diffusion as both exponents in the figure are equal, while maximum anisotropy is achieved when $\delta=1$. This latter case for which the maximum value of $\sigma$ is achieved and corresponds to the $\xi_1$ direction, the diffusion enhancement is attributable to the well-known Taylor dispersion mechanism~\cite{taylor1953dispersion}. The relatively smaller diffusion enhancement when $\delta=0$ may be explained by the fact that although convective transport of a scalar can be quick within a given eddy, the transport to an adjacent eddy is still predominantly controlled by the slow molecular diffusion.

Using in~\eqref{LeS0} the exponents computed in Fig.~\ref{fig:sig} , we can determine the asymptotic behaviours of the effective Lewis number and the effective burning speed ratio. In the case of square vortices ($\delta=0$), these are given by
\begin{equation}
    \Lew_{\rm{eff}} \simeq \Lew^{1/2}, \qquad  \qquad  \frac{S_T}{S_L} \sim \left(\frac{\Pec}{\Lew}\right)^{1/4} = \frac{1}{\Lew^{1/4}} \left(\frac{U}{S_L}\right)^{1/4} \left(\frac{l_{\mathrm{cell}}}{\delta_L}\right)^{1/4}, \label{sq}
\end{equation}
irrespective of the direction of flame propagation due to the isotropy of the effective diffusion process. As for the case of unidirectional shear ($\delta=1$), we have
\begin{equation}
    \Lew_{\rm{eff}} \simeq \frac{1}{\Lew}, \qquad  \qquad \frac{S_T}{S_L} \sim \frac{\Pec}{\Lew} = \frac{1}{\Lew}\frac{U}{S_L} \frac{l_{\mathrm{cell}}}{\delta_L}  \label{sh}
\end{equation}
when the flame propagates in the direction of the shear flow ($\xi_1$-direction). Of course, if we consider  flame propagation in the $\xi_2$-direction where there is no diffusion enhancement, then 
\begin{equation}
    \Lew_{\rm{eff}} = \Lew, \qquad  \qquad \frac{S_T}{S_L} =1.  \nonumber
\end{equation}
The behaviours of other values of $\delta$, lie between the above two limiting cases $\delta=0$ and $\delta=1$ just considered.  It should be noted that the $\frac{1}{4}$th power dependence of $S_T/S_L$ on $\Pec$ given in~\eqref{sq} for the case of square vortices was first identified by Audoly \textit{et.al.}~\cite{audoly2000reaction} and later confirmed in~\cite{kagan2002activation,abel2002front,vladimirova2003flame}; see also~\cite[pp.186-187]{clavin2016combustion}. On the other hand, the dependence on $\Pec$ and $\Lew$ in~\eqref{sh} for unidirectional shear flows was first reported in~\cite{daou2018taylor}.

\subsection{Potential implications for turbulent premixed combustion}
\label{sec:turb}

It is worth comparing the asymptotic behaviours~\eqref{sq}-\eqref{sh} with the corresponding trend in premixed turbulent combustion in the distributed reaction zone regime. We begin by comparing our results with the burning speed formula  
\begin{equation}
    \frac{S_T}{S_L} \sim \frac{\sqrt{\Rey}}{\Lew} \sim \frac{\Rey_\lambda}{\Lew} \label{exp}
\end{equation}
reported in the recent experimental study~\cite{cai2022turbulent} on highly turbulent jet flames; here $\Rey$ is the turbulent Reynolds number and $\Rey_\lambda$ is the Taylor-scale Reynolds number. Now according to Damköhler’s second hypothesis~\cite{damkohler1940einfluss}, the effect of small scale turbulence is to enhance the effective diffusion coefficients and hence the effective burning speed $S_T$, without altering the flame structure. Therefore, as we argued in~\cite{rajamanickam2023thick}, we may write  
\begin{equation}
    \frac{S_T}{S_L} = \sqrt{\frac{\Lew_{\rm{tur}} D_{\rm{T,tur}}}{\Lew\, D_T}}  \label{Dam}
\end{equation}
which follows from using formula~\eqref{zel}; here $\Lew_{\rm{tur}}=D_{T,\rm{tur}}/D_{F,\rm{tur}}$ is the turbulent Lewis number and $D_{T,\rm{tur}}$ and $D_{F,\rm{tur}}$ are the turbulent (or effective) thermal diffusivity  and fuel diffusion coefficient.

Comparing the last two relations, we find
\begin{equation}
    \Lew_{\rm{tur}}= \frac{1}{\Lew} \qquad \text{and} \qquad \frac{D_{T,\rm{tur}}}{D_T}\sim  \Rey. \label{Letur}
\end{equation}
The Lewis number dependence of the turbulent burning speed~\eqref{exp} and the turbulent Lewis number~\eqref{Letur} appear to be in better agreement with the predictions of unidirectional shear flow~\eqref{sh} than the square vortices~\eqref{sq}. This observation is somewhat surprising as the shear flow lacks more the isotropic aspect of turbulent diffusion coefficients than the cellular flows. Furthermore, we note that it is difficult to reconcile the dependence on the Reynolds number between the turbulent and the laminar flow cases. Yet, it is interesting to note that formulas~\eqref{exp} with~\eqref{sh} are in good agreement if $\Pec$ is identified with the  Reynolds number $\Rey_\lambda$ based on the Taylor microscale (rather than $\Rey)$.

It is also instructive to examine the dependence of the effective burning speed $S_T$, rather than $S_T/S_L$, on the Lewis number $\Lew$. In particular, since $S_L\propto \sqrt{\Lew}$, we have
\begin{equation}
    S_T \propto \Lew^{(1-\sigma)/2}
\end{equation}
or equivalently $S_T \propto \sqrt{\Lew_{\mathrm{eff}}}$. From the above relation, we can conclude that the effective burning speed $S_T$ decreases with increasing Lewis number only when $\sigma>1$. This trend for $S_T$ is also observed in turbulent cases, see e.g. figure 3 in~\cite{lipatnikov2005molecular}, which can therefore be explained, in part, by flow-enhanced diffusion.

Irrespective of the complications associated with turbulent combustion, our study highlights a physically important result. Specifically, the study shows that the flow plays a crucial part in determining the effective Lewis number, leading to surprising results such as $\Lew_{\mathrm{eff}}\simeq 1/\Lew$ for parallel flows~\eqref{sh} and $\Lew_{\mathrm{eff}}\simeq \Lew^{1/2}$ for square vortices~\eqref{sq} at large values of $\Pec$. Such results can provide explanations for unexpected flame behaviours in turbulent or complex flow fields. An example of such unexpected behaviours is the experimental observation reported in~\cite{wu2014facilitated,saha2019competing} that turbulence appears to facilitate ignition in $\Lew>1$ mixtures and to inhibit it in $\Lew<1$ mixtures. Partial explanation to this observation may be provided by the dependence of $\Lew_{\mathrm{eff}}$ on the flow field emphasized herein.

\section{Results for Childress-Soward flows with arbitrary Peclet numbers}
\label{sec:arbPe}

In the previous section, we have explored the asymptotic behaviours of the burning speed for large values of the Peclet number. Here we shall present illustrative results for arbitrary values of $\Pec$ in the case of Childress-Soward flows~\eqref{child}. We first consider the unity Lewis number case for which formula~\eqref{S0} implies that
\begin{equation}
    \frac{S_T}{S_L}\simeq \sqrt{\mathcal{D}_{T,11}'} \quad \text{or} \quad \frac{S_T}{S_L}\simeq \sqrt{\mathcal{D}_{T,22}'} \label{stslle1}
\end{equation}
depending on whether the flame propagates in the $\xi_1$-direction or the $\xi_2$-direction.

\begin{figure}[h!]
\includegraphics[width=0.49\textwidth]{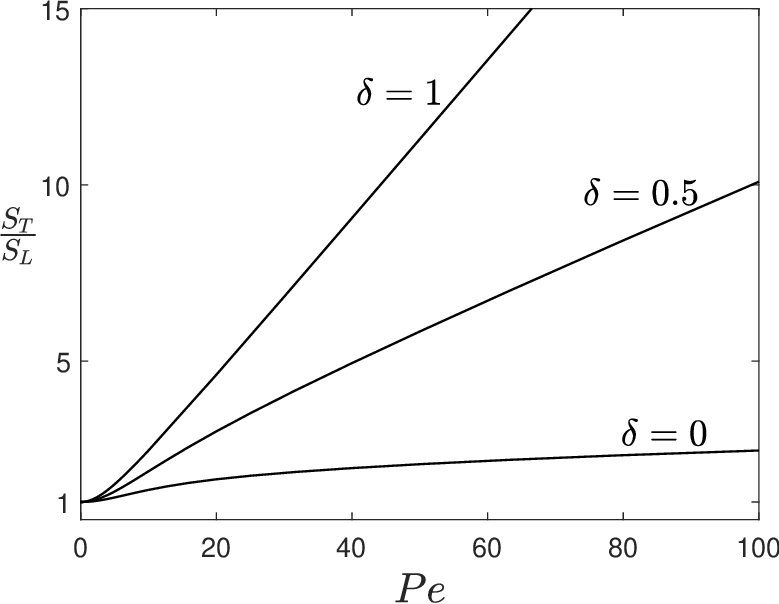}
\includegraphics[width=0.49\textwidth]{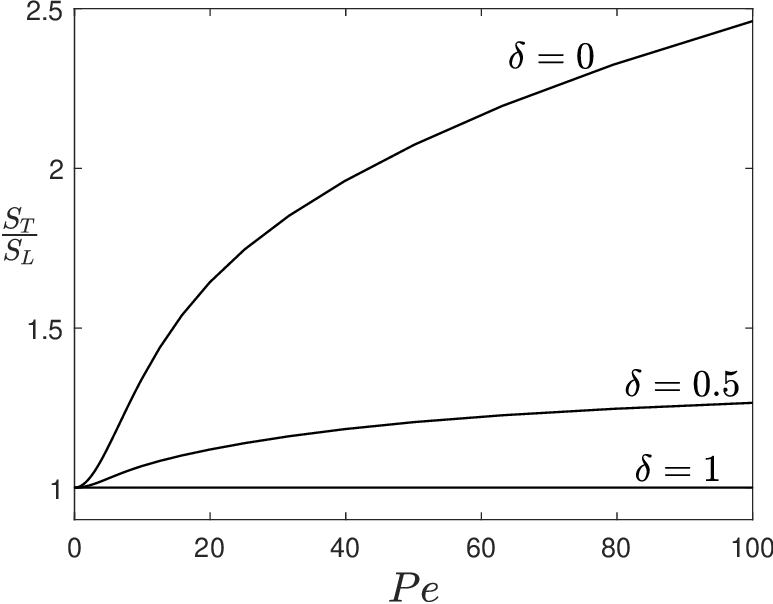}
\caption{The effective burning speed ratio $S_T/S_L$ vs. the Peclet number $\Pec$ for the Childress-Soward flows~\eqref{child} with $\Lew=1$. The left figure corresponds to flame propagation along the $\xi_1$-axis and the right figure to the $\xi_2$-axis.} \label{fig:stsl}
\end{figure}

Figure~\ref{fig:stsl} is generated by computing the effective diffusion coefficients in~\eqref{DF}-\eqref{DT} for different values of $\Pec$ and substituting into~\eqref{stslle1}. The curves of $S_T/S_L$ reveal a quadratic dependence on $\Pec$ for small values of $\Pec$, whereas at large values they approach the asymptotic behaviour identified in the previous section. In particular, it is worth noting at large values of $\Pec$ the linear behaviour for $\delta=1$ and the sublinear behaviour when $\delta<1$ which exhibits a bending effect of the curve $S_T/S_L$ vs. $\Pec$.

It is worth noting that the curve for the periodic shear flow ($\delta=1$) with flame propagation along $\xi_1$-direction may be compared with the corresponding curves reported for Poiseuille flows. Specifically, our findings are consistent with the curves in figures 5 and 6 of~\cite{daou2001flame} and figure 8 of~\cite{short2009asymptotic}. In the case of cellular flows ($\delta=0$), the findings are found to be consistent with the result exhibited in Fig.~4.32 of~\cite{al2009influence}.

\begin{figure}[h!]
\includegraphics[width=0.49\textwidth]{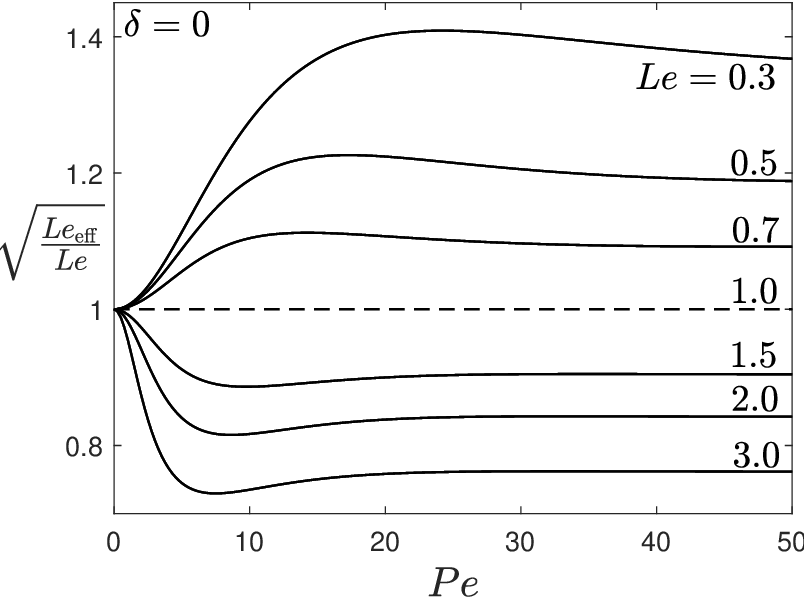}
\includegraphics[width=0.49\textwidth]{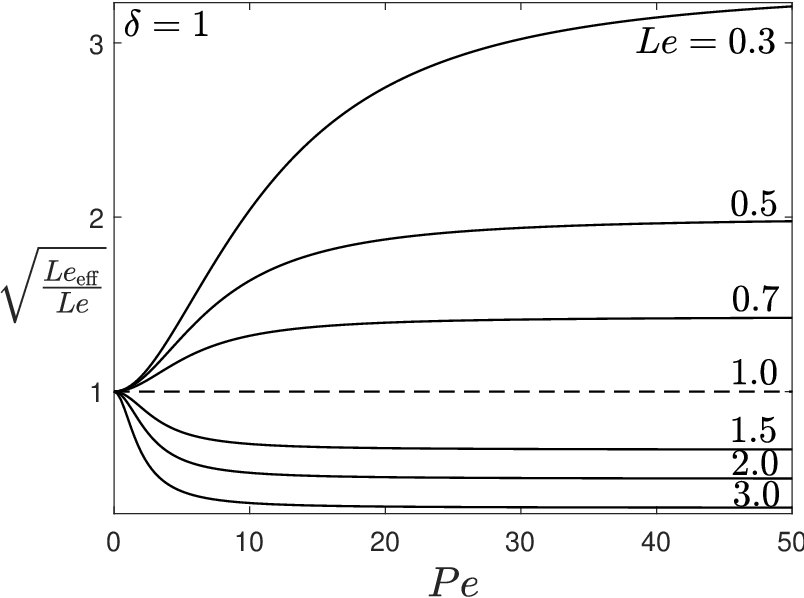}
\caption{The factor $\sqrt{\Lew_{\mathrm{eff}}/\Lew}$ (\eqref{enh}) as a function of $\Pec$ for selected values of $\Lew$. The left figure corresponds to the cellular flow ($\delta=0$) and the right figure to the shear flow ($\delta=1$). In both cases, flame is assumed to propagate in the $\xi_1$-direction.} \label{fig:Lef}
\end{figure}

We now examine the influence of non-unity Lewis numbers. To this end, we note that the burning speed ratio $S_T/S_L$ for non-unity Lewis numbers can be obtained according to formula~\eqref{S0} by multiplying the corresponding ratio $S_T/S_L$ for unity Lewis numbers plotted in Fig.~\ref{fig:stsl} by the factor $\sqrt{\Lew_{\mathrm{eff}}/\Lew}$. In other words, the factor
\begin{equation}
    \sqrt{\frac{\Lew_{\mathrm{eff}}}{\Lew}}=\frac{S_T/S_L}{(S_T/S_L)|_{\Lew=1}} \label{enh}
\end{equation}
is a convenient way to quantity the departure of the scaled burning speed from its unit Lewis number value. This factor is computed using~\eqref{DF}-\eqref{DT} and~\eqref{S0}  and is plotted as a function of $\Pec$ in Fig.~\ref{fig:Lef} for selected values of $\Lew$. All curves in this figure are found to exhibit a quadratic behaviour near $\Pec=0$ and asymptote to the value $1/\Lew^{\sigma/2}$ for large $\Pec$.

\section{Effects of thermal expansion and heat loss}
\label{sec:th}

Influence of thermal expansion and heat loss can be taken into account in a straightforward manner because the primary change that encountered here is the enhancement of diffusion coefficients. First, let us address thermal expansion effects. It is clear that density variations associated with thermal expansion due to heat release must be in the first approximation a function of $\vect X$ given by $\rho_0(\vect X)$, as has been shown in the related confined geometry problems~\cite{pearce2014taylor,daou2018taylor,rajamanickam2022effects,rajamanickam2023thick}. This means that density is practically constant on the small-scale variables $(\mathbf{\xi},\tau)$. On account of the density variation on the large scale, the effective diffusion coefficients~\eqref{DF} and~\eqref{DT} will depend on $\rho_0(\vect X)$; see e.g. formulas (21)-(22) in~\cite{daou2018taylor}. It follows that the required change in our asymptotic formulas~\eqref{S0} is that the the diffusion coefficients need simply to be evaluated at the burnt gas temperature. This implies, for example, that formula~\eqref{sigma} need to be replaced with
\begin{equation}
    \mathcal{D}_{F,11}'-1 \sim (\mathcal{P}\Lew)^\sigma \quad \text{where} \quad \mathcal{P}=\Pec(1-\alpha)
\end{equation}
is the Peclet number involving the gas expansion parameter $\alpha$ defined in~\eqref{zel}. \blue{The reader is referred to~\cite{rajamanickam2023thick} for an analysis that incorporates the thermal expansion in a related simpler problem.}

Turning now to the effect of heat loss, let us assume that the heat loss rate per unit volume may be written as $\rho c_p K (T-T_u)$, where $K^{-1}$ is a characteristic cooling time. The scaling of $K$ for flame quenching is given by $K\delta_L^2/D_T \sim 1/\beta$. Thus, we can introduce the parameter $\kappa= \beta K\delta_L^2/D_T$, which introduces on the right-hand side a term $-\ep^2\kappa \theta /\beta$ in~\eqref{theta} and correspondingly a term $-\kappa\Theta_0/\beta$ in~\eqref{T0}. The classical asymptotic result~\cite{joulin1979linear} for the burning speed with account taken of diffusion enhancement then becomes
\begin{equation}
    S_{0}^2 \ln \frac{S_0}{S_{0,ad}} = - \kappa \mathcal{D}_{T,11}'
\end{equation}
where $S_{0,ad}$ is the adiabatic flame speed given by~\eqref{S0}. The burning speed $S_0$ exists for $\kappa\leq \kappa_{ext}$, where 
\begin{align}
    \kappa_{ext} = \frac{S_{0,ad}^2}{2e\mathcal{D}_{T,11}'}=\frac{\Lew_{\mathrm{eff}}}{2e\,\Lew}.
\end{align}
For large Peclet numbers with $\sigma>0$, this formula simplifies to
\begin{equation}
    \kappa_{ext} = \frac{1}{2e\, \Lew^{\sigma}}
\end{equation}
which indicates that for $\Lew<1$, the extinction limit is enlarged by a factor $1/\Lew^\sigma$ in the presence of the periodic flow, whereas it is diminished by the same amount for $\Lew>1$. This effect of diffusion enhancement on flame quenching is greatest when $\sigma=2$, as identified for Taylor-dispersion controlled flames in~\cite{rajamanickam2023thick}.

\section{Possible extensions of the study}
\label{sec:fut}

Although, the results derived herein pertain to zero-mean, small-scale periodic flows, they are also applicable if the flows have a small non-zero mean which varies on the large scale. For example, the formulas for the generalized diffusion matrices derived in~\eqref{DF}-\eqref{DT} are still applicable for the flow field $\vect v(\vect x/l_{\mathrm{cell}}, t/t_{\mathrm{cell}})/U + \ep \vect V(\vect x/\delta_L,tS_L/\delta_L)$ where $\vect V$ denotes the large-scale weak mean flow. The convection velocity $\ep {S}\vect n\approx \ep S_0\vect n$ in~\eqref{yF}-\eqref{theta}, emerging in the flame-fixed frame, is itself a weak mean flow, albeit a constant one at leading order. Excluding some peculiar cases discussed in~\cite[$\S$~2.1.3.1]{majda1999simplified}, there appears to be no theoretical development for the large-scale mean flow of arbitrary magnitude. The latter problem is also of considerable interest for future investigations.

To close this section, we note that findings of the present paper are applicable, strictly speaking, when $\delta_L/\beta\gg l_{\rm{cell}}$. When the flow scale $l_{\rm{cell}}$ is of the order of, or slightly larger than, the reaction zone thickness $\delta_L/\beta$, further progress can be made provided $l_{\rm{cell}}\ll \delta_L$. This can be done by carrying out an asymptotic analysis in the distinguished limit $\delta_L/\beta\sim l_{\rm{cell}}$, as done in~\cite{rajamanickam2023thick} for unidirectional flows. In this distinguished limit, the theoretical approach developed here can be applied to the preheat and post flame zones, but for the reaction zone, a convective-diffusive-reactive inner problem is obtained. \blue{Specifically, consideration of this distinguished limit shows that the first corrections to the leading-order solutions obtained here are of order $\beta\ep=\beta l_{\rm{cell}}/\delta_L$.}


\section{Concluding remarks}
\label{sec:con}
In this paper, we have carried out an asymptotic analysis of the propagation of a thick flame in small-scale, zero-mean, spatio-temporal periodic flows. Using activation energy asymptotics and homogenization theory, formulas~\eqref{S0} for the effective Lewis number $\Lew_{\mathrm{eff}}$ and the effective burning speed ratio $S_T/S_L$ have been derived. The formulas quantify the dependence of the propagation and structure of the flame on the flow via flow enhanced diffusion. In particular, when the flow Peclet number $\Pec$ is large, the enhanced  fuel diffusion coefficient and the enhanced thermal diffusivity are found to grow like $(\Pec\Lew)^\sigma$ and $\Pec^\sigma$, respectively, where $\sigma\leq 2$ is a constant that depends on the flow and the direction of flame propagation. Consequently, this leads to $\Lew_{\mathrm{eff}}\simeq \Lew^{1-\sigma}$ and $S_T/S_L \sim (\Pec/\Lew)^{\sigma/2}$; the maximal diffusion enhancement ($\sigma=2$) is achieved for steady, unidirectional flows. The result also indicates that the effective burning speed $S_T\propto \Lew^{(1-\sigma)/2}$ increases with decreasing Lewis numbers only when $\sigma>1$. 

We have also briefly addressed the effect of heat loss on flame propagation and quenching in a flow field in $\S$\ref{sec:th}. In particular, it is worth noting that the quenching limit due to heat loss is increased by the factor $1/\Lew^\sigma$ due to the presence of the flow. That is to say, small-scale flows increase the quenching limit of subunity Lewis-number mixtures and decrease it for mixtures with $\Lew>1$.

Finally, the potential implications of the findings to better understand turbulent combustion have been summarized in $\S$\ref{sec:turb}. A particular aspect which has been emphasised in our study is the dependence of the flame characteristics on the Lewis number in the presence of high-intensity, small-scale flows. This dependence has been shown to intimately depend on the flow through Taylor-dispersion like flow-enhanced diffusion, rather than merely on the conventional molecular diffusion coupled with curvature effects. We have also argued that the flow-dependent effective Lewis number identified herein may be useful to explain the peculiar feature that turbulence appears to facilitate ignition in $\Lew>1$ mixtures and to inhibit it in $\Lew<1$ mixtures, observed in experiments on ignition in a turbulent reactive flow~\cite{wu2014facilitated,saha2019competing}.


\section*{Acknowledgements}
This research was supported by the UK EPSRC through grant EP/V004840/1.

\bibliographystyle{elsarticle-num}

\bibliography{sample}

\begin{thebibliography}{10}
\expandafter\ifx\csname url\endcsname\relax
  \def\url#1{\texttt{#1}}\fi
\expandafter\ifx\csname urlprefix\endcsname\relax\def\urlprefix{URL }\fi
\expandafter\ifx\csname href\endcsname\relax
  \def\href#1#2{#2} \def\path#1{#1}\fi

\bibitem{ashurst1991short}
W.~T. Ashurst, G.~I. Sivashinsky, On flame propagation through periodic flow
  fields, Combust. Sci. Technol. 80 (1991) 159--164.

\bibitem{berestycki1991flame}
H.~Berestycki, G.~I. Sivashinsky, Flame extinction by periodic flow field, SIAM
  J. Appl. Math. 51 (1991) 344--350.

\bibitem{brailovsky1995extinction}
I.~Brailovsky, G.~I. Sivashinsky, Extinction of a nonadiabatic flame
  propagating through spatially periodic shear flow, Phys. Rev. E 51 (1995)
  1172.

\bibitem{audoly2000reaction}
B.~Audoly, H.~Berestycki, Y.~Pomeau, R{\'e}action diffusion en {\'e}coulement
  stationnaire rapide, C. R. Acad. Sci. Series IIB-Mechanics-Physics-Astronomy
  328 (2000) 255--262.

\bibitem{aldredge1996premixed}
R.~C. Aldredge, Premixed flame propagation in a high-intensity, large-scale
  vortical flow, Combust. Flame 106 (1996) 29--40.

\bibitem{kagan1998flame}
L.~Kagan, G.~I. Sivashinsky, G.~Makhviladze, On flame extinction by a spatially
  periodic shear flow, Combust. Theory Model. 2 (1998) 399.

\bibitem{kagan2000flame}
L.~Kagan, G.~I. Sivashinsky, Flame propagation and extinction in large-scale
  vortical flows, Combust. Flame 120 (2000) 222--232.

\bibitem{kagan2002activation}
L.~Kagan, P.~D. Ronney, G.~I. Sivashinsky, Activation energy effect on flame
  propagation in large-scale vortical flows, Combust. Theory Model. 6 (2002)
  479.

\bibitem{kagan2005effect}
L.~Kagan, G.~I. Sivashinsky, Effect of {L}ewis number on flame propagation
  through vortical flows, Combust. Flame 142 (2005) 235--240.

\bibitem{kadowaki2009dynamic}
S.~Kadowaki, H.~Kobayashi, Dynamic behavior of premixed flames propagating in
  non-uniform velocity fields—{A}ssessment of intrinsic instability in
  turbulent combustion—, Trans. Jpn. Soc. Aeronaut. Space Sci. 51 (2009)
  244--251.

\bibitem{kagan2020flame}
L.~Kagan, G.~I. Sivashinsky, Flame propagation through a spatially periodic
  shear flow: Extinction vs. transition to detonation, Combust. Flame 218
  (2020) 98--100.

\bibitem{feng2021influence}
R.~Feng, A.~Gruber, J.~H. Chen, D.~M. Valiev, Influence of gas expansion on the
  propagation of a premixed flame in a spatially periodic shear flow, Combust.
  Flame 227 (2021) 421--427.

\bibitem{shy1992experimental}
S.~S. Shy, P.~D. Ronney, S.~G. Buckley, V.~Yakhot, Experimental simulation of
  premixed turbulent combustion using aqueous autocatalytic reactions, in:
  Symp. (Int.) Combust., Vol.~24, Elsevier, 1992, pp. 543--551.

\bibitem{zhu1994simulation}
J.~Zhu, P.~D. Ronney, Simulation of front propagation at large non-dimensional
  flow disturbance intensities, Combust. Sci. Technol. 100 (1994) 183--201.

\bibitem{vaezi2000laminar}
V.~Vaezi, R.~C. Aldredge, Laminar-flame instabilities in a {T}aylor-{C}ouette
  combustor, Combust. Flame 121 (2000) 356--366.

\bibitem{peters1941turbulent}
N.~Peters, Turbulent combustion, Cambridge University Press, Cambridge, UK,
  2000.

\bibitem{lipatnikov2005molecular}
A.~N. Lipatnikov, J.~Chomiak, Molecular transport effects on turbulent flame
  propagation and structure, Prog. Energy Combust. Sci. 31 (2005) 1--73.

\bibitem{williams2000progress}
F.~A. Williams, Progress in knowledge of flamelet structure and extinction,
  Prog. Energy Combust. Sci. 26 (2000) 657--682.

\bibitem{rajamanickam2023thick}
P.~Rajamanickam, J.~Daou, A thick reaction zone model for premixed flames in
  two-dimensional channels, Combust. Theory Model. 27 (2023) 487--507.

\bibitem{aspden2011turbulence}
A.~J. Aspden, M.~Day, J.~B. Bell, Turbulence--flame interactions in lean
  premixed hydrogen: transition to the distributed burning regime, J. Fluid
  Mech. 680 (2011) 287--320.

\bibitem{lapointe2015differential}
S.~Lapointe, B.~Savard, G.~Blanquart, Differential diffusion effects,
  distributed burning, and local extinctions in high karlovitz premixed flames,
  Combust. Flame 162 (2015) 3341--3355.

\bibitem{aspden2019towards}
A.~J. Aspden, M.~S. Day, J.~B. Bell, Towards the distributed burning regime in
  turbulent premixed flames, J. Fluid Mech. 871 (2019) 1--21.

\bibitem{rieth2022enhanced}
M.~Rieth, A.~Gruber, F.~A. Williams, J.~H. Chen, Enhanced burning rates in
  hydrogen-enriched turbulent premixed flames by diffusion of molecular and
  atomic hydrogen, Combust. Flame 239 (2022) 111740.

\bibitem{lee2021influence}
H.~C. Lee, P.~Dai, M.~Wan, A.~N. Lipatnikov, Influence of molecular transport
  on burning rate and conditioned species concentrations in highly turbulent
  premixed flames, J. Fluid Mech. 928.

\bibitem{lee2022lewis}
H.~C. Lee, P.~Dai, M.~Wan, A.~N. Lipatnikov, Lewis number and preferential
  diffusion effects in lean hydrogen--air highly turbulent flames, Phys. Fluids
  34 (2022) 035131.

\bibitem{cai2022turbulent}
X.~Cai, Q.~Fan, X.-S. Bai, J.~Wang, M.~Zhang, Z.~Huang, M.~Alden, Z.~Li,
  Turbulent burning velocity and its related statistics of ammonia-hydrogen-air
  jet flames at high {K}arlovitz number: Effect of differential diffusion,
  Proc. Combust. Inst.

\bibitem{daou2002thick}
J.~Daou, J.~W. Dold, M.~Matalon, The thick flame asymptotic limit and
  {D}amk{\"o}hler's hypothesis, Combust. Theory Model. 6 (2002) 141.

\bibitem{majda1999simplified}
A.~J. Majda, P.~R. Kramer, Simplified models for turbulent diffusion: theory,
  numerical modelling, and physical phenomena, Phys. Rep. 314 (1999) 237--574.

\bibitem{daou2018taylor}
J.~Daou, P.~Pearce, F.~Al-Malki, Taylor dispersion in premixed combustion:
  Questions from turbulent combustion answered for laminar flames, Phys. Rev.
  Fluids 3 (2018) 023201.

\bibitem{zeldovich1982exact}
{\relax Ya}.~B. Zeldovich, An exact solution to the problem of diffusion in a
  periodic velocity field and turbulent diffusion, Dokl. Akad. Nauk SSSR 266
  (1982) 821--826.

\bibitem{watson1983diffusion}
E.~J. Watson, Diffusion in oscillatory pipe flow, J. Fluid Mech. 133 (1983)
  233--244.

\bibitem{childress1989scalar}
S.~Childress, A.~M. Soward, Scalar transport and alpha-effect for a family of
  cat's-eye flows, J. Fluid Mech. 205 (1989) 99--133.

\bibitem{childress1979alpha}
S.~Childress, Alpha-effect in flux ropes and sheets, Phys. Earth Planet. Inter.
  20 (1979) 172--180.

\bibitem{fannjiang1994convection}
A.~Fannjiang, G.~Papanicolaou, Convection enhanced diffusion for periodic
  flows, SIAM J. Appl. Math. 54 (1994) 333--408.

\bibitem{taylor1953dispersion}
G.~I. Taylor, Dispersion of soluble matter in solvent flowing slowly through a
  tube, Proc. Roy. Soc. Lond. Series A. Math. Phys. Sci. 219 (1953) 186--203.

\bibitem{abel2002front}
M.~Abel, M.~Cencini, D.~Vergni, A.~Vulpiani, Front speed enhancement in
  cellular flows, Chaos 12 (2002) 481--488.

\bibitem{vladimirova2003flame}
N.~Vladimirova, P.~Constantin, A.~Kiselev, O.~Ruchayskiy, L.~Ryzhik, Flame
  enhancement and quenching in fluid flows, Combust. Theory Model. 7 (2003)
  487.

\bibitem{clavin2016combustion}
P.~Clavin, G.~Searby, Combustion waves and fronts in flows: Flames, shocks,
  detonations, ablation fronts and explosion of stars, Cambridge University
  Press, Cambridge, UK, 2016.

\bibitem{damkohler1940einfluss}
G.~Damk{\"o}hler, Der einfluss der turbulenz auf die flammengeschwindigkeit in
  gasgemischen, Z. Elektrochem. Angew. Phys. Chem. 46 (1940) 601--626.

\bibitem{wu2014facilitated}
F.~Wu, A.~Saha, S.~Chaudhuri, C.~K. Law, Facilitated ignition in turbulence
  through differential diffusion, Phys. Rev. Lett. 113 (2014) 024503.

\bibitem{saha2019competing}
A.~Saha, S.~Yang, C.~K. Law, On the competing roles of turbulence and
  differential diffusion in facilitated ignition, Proc. Combust. Inst. 37
  (2019) 2383--2390.

\bibitem{daou2001flame}
J.~Daou, M.~Matalon, Flame propagation in poiseuille flow under adiabatic
  conditions, Combust. Flame 124 (2001) 337--349.

\bibitem{short2009asymptotic}
M.~Short, D.~A. Kessler, Asymptotic and numerical study of variable-density
  premixed flame propagation in a narrow channel, J. Fluid Mech. 638 (2009)
  305--337.

\bibitem{al2009influence}
F.~A. Al-Malki, Influence of flow on the propagation of premixed and partially
  premixed flames, Ph.D. thesis, University of Manchester (2009).

\bibitem{pearce2014taylor}
P.~Pearce, J.~Daou, Taylor dispersion and thermal expansion effects on flame
  propagation in a narrow channel, J. Fluid Mech. 754 (2014) 161--183.

\bibitem{rajamanickam2022effects}
P.~Rajamanickam, A.~D. Weiss, Effects of thermal expansion on {T}aylor
  dispersion-controlled diffusion flames, Combust. Theory Model. 26 (2022)
  50--66.

\bibitem{joulin1979linear}
G.~Joulin, P.~Clavin, Linear stability analysis of nonadiabatic flames:
  {D}iffusional-thermal model, Combust. Flame 35 (1979) 139--153.

\end{thebibliography}

\end{document}